\documentclass[fleqn,usenatbib]{mnras}

\usepackage{newtxtext,newtxmath}

\usepackage[T1]{fontenc}
\usepackage{ae,aecompl}

\usepackage{graphicx}	
\usepackage{amsmath}	
\usepackage{amssymb}	
\usepackage[utf8]{inputenc}

\title[Properties of the ULX X-3 in NGC 4258]{The ULX Source X-3 in NGC 4258: \\
A Search for Its X-ray and Optical Properties}

\author[A. Akyuz et al.]{A. Akyuz,$^{1,2}$\thanks{E-mail:aakyuz@cu.edu.tr} S. Avdan,$^{2}$ S. Allak,$^{2,3}$ N. Aksaker,$^{2,4}$ İ. Akkaya Oralhan$^{5}$ \and Ş. Balman$^{6,7}$
\\
$^1$Department of Physics, University of Çukurova, 01330, Adana, Turkey\\
$^2$Space Science and Solar Energy Research and Application Center (UZAYMER), University of Çukurova, 01330, Adana, Turkey\\
$^3$Department of Physics, University of Çanakkale Onsekiz Mart, 17020, Çanakkale, Turkey\\
$^4$Adana Organised Industrial Zones Vocational School of Technical Science, University of Çukurova, 01410, Adana, Turkey\\
$^5$Department of Astronomy and Space Sciences, Erciyes University, 38039, Kayseri, Turkey\\
$^6$Department of Astronomy and Space Sciences, Istanbul University, 34119, Istanbul, Turkey\\
$^7$Faculty of Engineering and Natural Sciences, Kadir Has University, Cibali, 34083, Istanbul, Turkey\\
}

\date{Accepted 2020 September 14. Received 2020 August 12; in original form 2020 March 22.}

\pubyear{2020}

\begin{document}
\label{firstpage}
\pagerange{\pageref{firstpage}--\pageref{lastpage}}
\maketitle

\begin{abstract}
We present the results of a search for the nature of ultraluminous X-ray source (ULX) X-3 in the nearby galaxy NGC 4258. We use archival data from {\it XMM-Newton}, {\it Chandra}, {\it NuSTAR} and {\it HST} observations. Total X-ray data analysed to find the model parameters of the system is indicative of a stellar mass black hole, $\sim$ 10 M$_{\sun}$, as the central compact object. Furthermore, analyses of the optical data from {\it HST} reveal two optical candidates with the 90 per cent confidence level of error radius of 0$\farcs$28. Assuming the optical emission is dominated by the donor star, both of these candidates are found to have spectral types that lie between B3$-$F1 with absolute magnitudes of $M_{\rm V}$ $\approx$ -6.4. Moreover, the age and mass estimates for the candidates are found to be of 10 and 18 Myr and of 13 and 20 M$_{\sun}$, respectively.
\end{abstract}




\section{Introduction}
\label{sec:Introduction}

Ultraluminous X-ray sources (ULXs) are off-nuclear point sources in a number of external galaxies. The X-ray luminosity of such a source is usually $L_{X}$ $\sim$ $10^{39}$ erg $s^{-1}$, exceeding the Eddington limit for a typical 10 M$_{\sun}$ stellar-remnant black hole (see the review by \citealp{2017ARA&A..55..303K}). Models of ULXs proposed to explain the high luminosities of ULXs require that a central accretor in the form of a black hole should exist as a driving engine. This proposed picture represent sources that may be either a stellar or an intermediate mass black hole. If a ULX hosts a stellar mass black hole (sMBH), the high luminosity is generated by super-Eddington accretion onto the stellar mass accretor \citep{2013MNRAS.432..506P, 2013MNRAS.435.1758S, 2014Natur.514..198M, 2015NatPh..11..551F}.
However, if a ULX hosts an intermediate mass black hole (IMBH), the high luminosity could then be explained by sub-Eddington accretion onto such sources \citep{2004IJMPD..13....1M, 2009Natur.460...73F, 2015MNRAS.448.1893M}.

However, recent discoveries proved that some ULXs do show coherent pulsations and they are classified as pulsating ULXs (PULXs). In this case, the luminosity is generated from a super-Eddington accretion onto a magnetized neutron star orbiting a stellar companion \citep{2014Natur.514..202B, 2017Sci...355..817I, 2016ApJ...831L..14F, 2017MNRAS.466L..48I, 2018MNRAS.476L..45C, 2019MNRAS.488L..35S, 2019arXiv190604791R}.
Hence, the debate is continuing about the type of the dominant ULX accretor \citep{2016MNRAS.458L..10K, 2017xru..conf..237W}.

Based on the high quality {\it XMM-Newton} and {\it Chandra} data, it became clear that the observed X-ray spectra of ULXs are different from those of Galactic black hole binaries (BHBs). The majority of ULX spectra revealed a curvature that is described by a cut-off usually at $\sim$ 3-7 keV, and mostly accompanied by a soft excess. The curvature is often interpreted either as an effect of a cold optically thick corona or emission from the inner regions of a geometrically thick accretion disk \citep{2006MNRAS.368..397S, 2007Ap&SS.311..203R, 2009MNRAS.397.1836G}. These spectral features suggest a new {\it ultraluminous} accretion state for ULXs where super-Eddington accretion flows onto a black hole or as recent pulsating ULXs show, onto a neutron star.
Broad-band {\it NuSTAR} observations have clarified that the curvature of the spectra of ULXs have a high energy cutoff extending above 10 keV. On the other hand, for the galactic BHBs, this spectral cutoff typically exceeds $\sim$60 keV. \citep{2013ApJ...778..163B, 2016MNRAS.460.4417L, 2017ApJ...836..113P, 2017ApJ...834...77F}.

Optical studies of ULXs also have important implications for understanding the nature of these sources and their environments. Especially, the optical identification of counterparts may give us valuable information on the mass and spectral type of the companion star and the origin of the optical emission. This emission is thought to originate either from the accretion disk by a reprocessing of X-rays in the outer disk regions and/or from the companion star. So far, single or multiple optical counterparts were identified for about 30 ULXs using ground-based telescopes and Hubble Space Telescope ({\it HST}) \citep{2005MNRAS.356...12S, 2011ApJ...737...81T, 2011ApJ...734...23G, 2012ApJ...745..123G, 2013ApJS..206...14G, 2016MNRAS.455L..91A, 2019ApJ...875...68A, 2019MNRAS.488.5935A}.

Spectra of optical counterparts usually allow us to construct radial velocity curves. However, optical spectra of ULXs generally lack traceable absorption lines (most likely due to their faintness (m$_{V}$ $\geq$ 21)), or have high-energy emission lines like HeII 4686 that do not seem to trace any clear orbital movement. Even in the case of P13 with absorption lines from the companion star are visible, they do not trace any orbital movement in the system. They are most likely due to X-ray (uneven) heating of the donor star \citep{2014Natur.514..198M}.

The absolute magnitudes and color indices of optical counterparts are found to lie in the intervals M$_{V}$ $=$ $-$3 to $-$8 and $B-V$ $=$ $-$6 to $+$0.4, respectively. ULX systems, most probably, contain OB supergiant donors. The blue color observed is thought to arise from the X-ray reprocessing in the accretion disk or from the donor star, or both
\citep{2010MNRAS.403L..69P, 2012ApJ...758...28J, 2018ApJ...854..176V}. However, some ULXs do contain red supergiant companions in the near-infrared band \citep{2014MNRAS.442.1054H, 2016MNRAS.459..771H, 2017MNRAS.469..671L}.

Most of ULXs are found to be located inside star forming regions and a few of them are reported to be powering the surrounding nebula by their radiation and/or outflows \citep{2003RMxAC..15..197P, 2007ApJ...668..124A, 2017ARA&A..55..303K}. There are also, many ULXs which are associated with young (5$-$20 Myr) star clusters helping to understand some of their subtle properties \citep{2005MNRAS.356...12S, 2011ApJ...734...23G, 2013MNRAS.432..506P,2016MNRAS.455L..91A, 2016ApJ...828..105A}.

\begin{figure*}
\begin{center}
\includegraphics[height=\columnwidth]{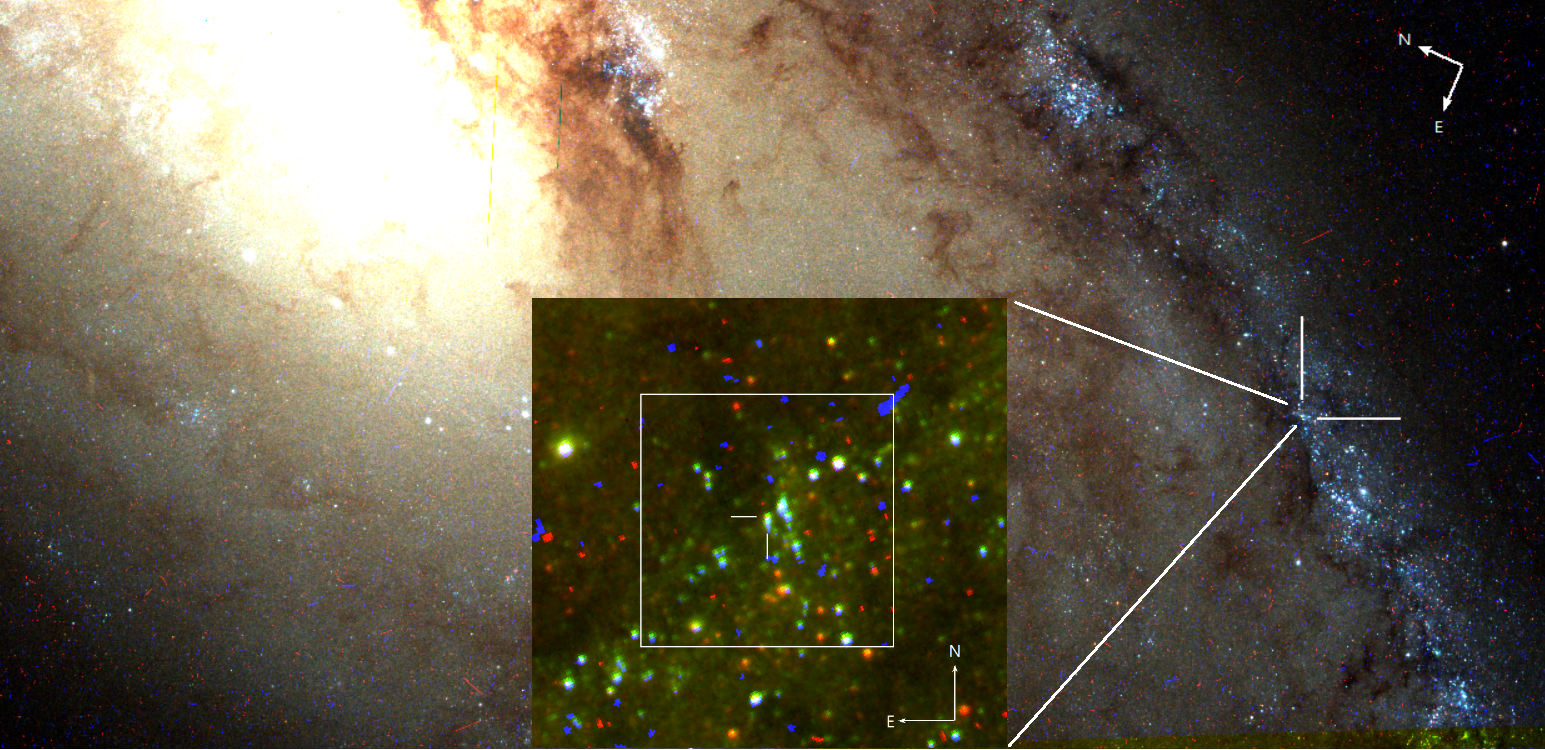}
\caption{The {\it HST}/ACS true color image of galaxy NGC 4258. Red, green and blue color represent F814W, F555W and F435W filters, respectively. X-3 position is marked with white bars. In the zoomed image, the 5\arcsec × 5\arcsec white box contains X-3 and field stars.}
\label{F:ngc4258}
\end{center}
\end{figure*}

In the present study, we searched for the X-ray spectral and temporal properties and also optical counterpart(s) of the ULX X-3 (hereafter X-3) in NGC 4258 using archival {\it XMM-Newton}, {\it Chandra}, {\it NuSTAR} and {\it HST} observations. This source is identified as a ULX with an X-ray luminosity of $L_{x}$=5.3 $\times$ $10^{39}$ erg $s^{-1}$ by \cite{2011ApJ...741...49S}.
NGC 4258 is classified as a Seyfert-type spiral galaxy at a distance 7.7 Mpc \citep{ 1991rc3..book.....D}. Our target source in NGC 4258 appears as a bright point-like source in the X-ray images. The source is located at the south of an arm at R.A.=$12^{h}$ $18^{m}$ $57^{s}.8$, Dec.=$+$ $47^{\circ}$ $16\arcmin$ $07\arcsec$ and $2\farcm1$ away from the galaxy center. The true color {\it HST} image of NGC 4258 is given in Fig. \ref{F:ngc4258} and the ULX is marked by lines.

The present introduction will be followed by description of observations and details of data analysis in Section \ref{sec:observation}. The discussion and conclusions of X-ray spectral analysis and optical properties of the X-3 are presented in Section \ref{sec:discussion}.

\section{Observations and Analysis Results}
\label{sec:observation}
\subsection{X-ray Observations}
\label{sssec:xray}
The source X-3 was observed by {\it XMM-Newton}, {\it Chandra} and {\it NuSTAR} satellites several times. The observations used in this study are listed in Table \ref{T:T1}.

The {\it XMM-Newton} data were analysed with standard Science Analysis Software ({\scshape sas}, v16.0.0) with well defined analysis steps. The events corresponding to single-double pixel (PATTERN $\leq$12) and single-multiple (PATTERN $\leq$4) with FLAG $=$ 0 were selected for EPIC MOS and PN cameras, respectively. The source and background spectra were extracted using $15\arcsec$ circular regions with {\scshape evselect} task in {\scshape sas}. Only XM7 data was affected by the background flaring. The last 3 ks data were removed from the observation prior to source and background extractions. Average of net count rate of the source is 0.06 counts s$^{-1}$ within the periods of flaring, while the value is 0.04 counts s$^{-1}$ outside the flaring episodes.

The {\it Chandra} data reduction was performed with Interactive Analysis of Observations ({\scshape ciao}) software (v4.9). The level 2 event files were obtained using {\scshape chandra\_repro} script in {\scshape ciao}. The source X-3 was located on the ACIS-S3 (back-illuminated) chip. The source and background photons were extracted with {\scshape specextract} task using $5\arcsec$ circular regions.

The {\it NuSTAR} data on the other hand, were analysed using nupipeline tool based on {\it NuSTAR} Data Analysis Software ({\scshape NuSTARDAS}, v1.7.1) within the {\scshape heasoft} software and calibration data CALDB version 20191219. The source and background photons were extracted using $30\arcsec$ circular regions. In {\it NuSTAR} data reductions, the background regions were extracted from source free regions close to X-3. Prior to fitting, the spectrum was grouped to have minimum of 35 counts per bin.

\begin{figure*}
\begin{center}
\includegraphics[angle=0,scale=0.35]{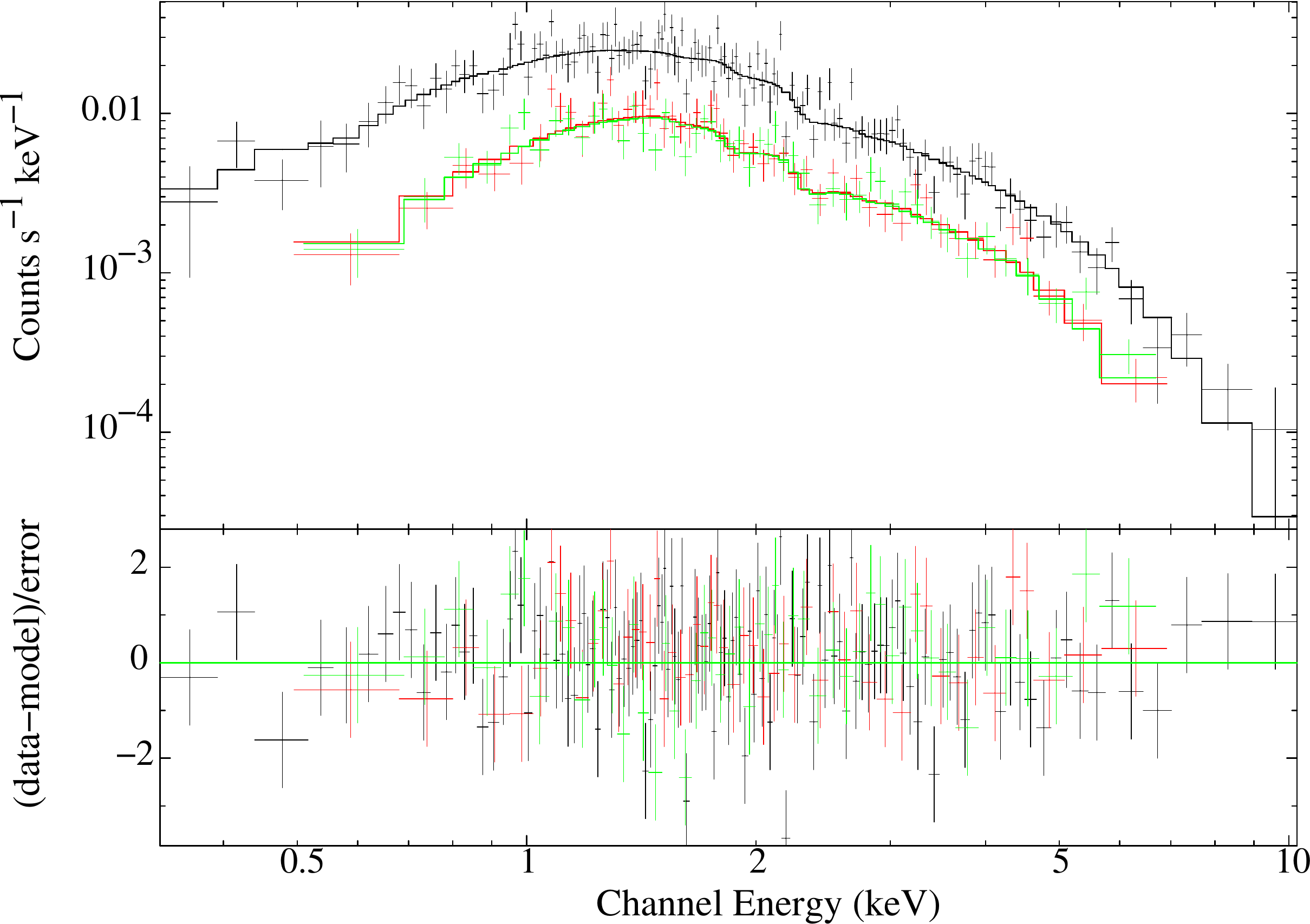}
\includegraphics[angle=0,scale=0.35]{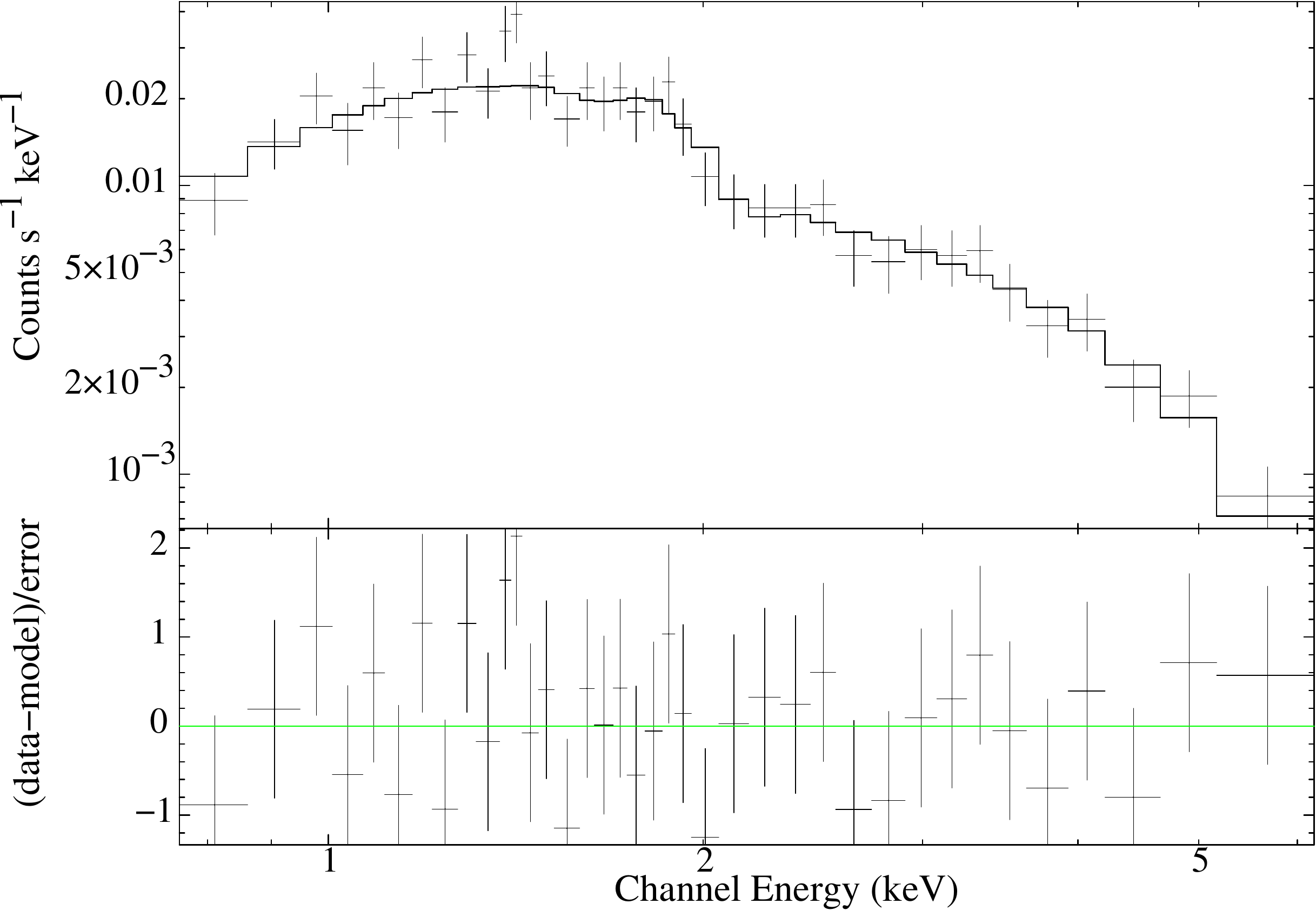}
\caption{Energy spectra of X-3 obtained using XM7 (left) and C1 (right) data. In {\it XMM-Newton} spectrum the black, green and red data points represent EPIC PN, MOS1 and MOS2, respectively. The {\it XMM-Newton} and {\it Chandra} spectra were fitted with diskbb model. Residuals of the fitting process are shown in the bottom panels.} 
\label{F:xrayspect}
\end{center}
\end{figure*}

The X-ray spectral fits were applied to the source spectra to interpret the origin of the X-ray emission and also to search for any spectral transitional behavior. {\it XMM-Newton} and {\it Chandra} spectral fits were performed using the {\scshape xspec} package (v12.9.1) in the 0.3$-$10 keV energy band. All spectra were grouped to have a minimum of 20 counts per bin before the fitting procedure. {\it XMM-Newton} EPIC PN and MOS data were fitted simultaneously by adding a constant model for instrumental calibration differences. The best-fit spectra were then obtained from the power-law (pl) and disk blackbody (diskbb) models, together with two absorption models (tbabs). One of the absorption models represented the line-of-sight column density which we kept fixed at the Galactic value (N$_{H}$ = $0.01\times10^{22}$ cm$^{-2}$; \citealp{1990ARA&A..28..215D}) and the other was left free to account for the intrinsic absorption. The unabsorbed flux values were calculated using CFLUX convolution model in the 0.3$-$10 keV energy band. The luminosity values were obtained by considering the adopted distance of NGC 4258 (7.7 Mpc). The best-fit model parameters for all observations are given in Table \ref{T:T2}. The energy spectra for XM7 and C1 data are given in Fig. \ref{F:xrayspect}. To investigate if flux variability occurred during the observations, the fitting process was repeated by fixing the intrinsic $N_{H}$ parameters to the average of the calculated values in the initial fitting. The calculated model parameters with this method are also given in Table \ref{T:T2} by denoting ``*'' and it is noted that the flux variability between, {\it XMM-Newton} observations is less than a factor of two.

\begin{figure}
\begin{center}
\includegraphics[scale=0.50]{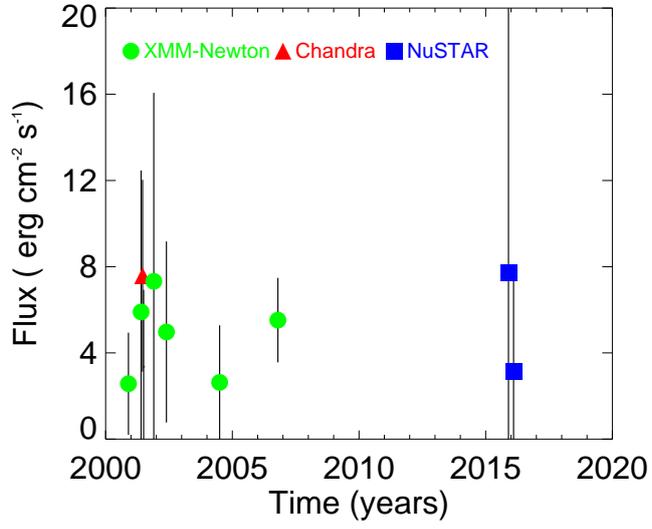}
\caption{Long-term light curve of X-3 obtained using all X-ray data. Circles (green), triangles (red) and squares (blue) represent {\it XMM-Newton}, {\it Chandra} and {\it NuSTAR} data, respectively. The fluxes were calculated between 3$-$10 keV energy range.}
\label{F:long-term}
\end{center}
\end{figure}

The long-term light curve of the source was constructed using the available X-ray data (Table \ref{T:T1}) to examine the flux variability. We used flux values obtained from pl model in 3$-$10 keV. No significant variability was visible and the source exhibits 2.8 factor difference between the lowest and the highest flux values. The long-term light curve of X-3 is given in Fig. \ref{F:long-term}

The spectral fitting was also carried out for {\it XMM-Newton}+{\it NuSTAR} spectra especially to investigate whether the cut-off is seen with {\it NuSTAR}, or if there is perhaps an extra component in the high-energy spectrum \citep{2013ApJ...778..163B, 2017ApJ...836..113P, 2018ApJ...869..111W, 2018ApJ...867..110B}.

\begin{figure*}
\begin{center}
\includegraphics[scale=0.55,angle=270]{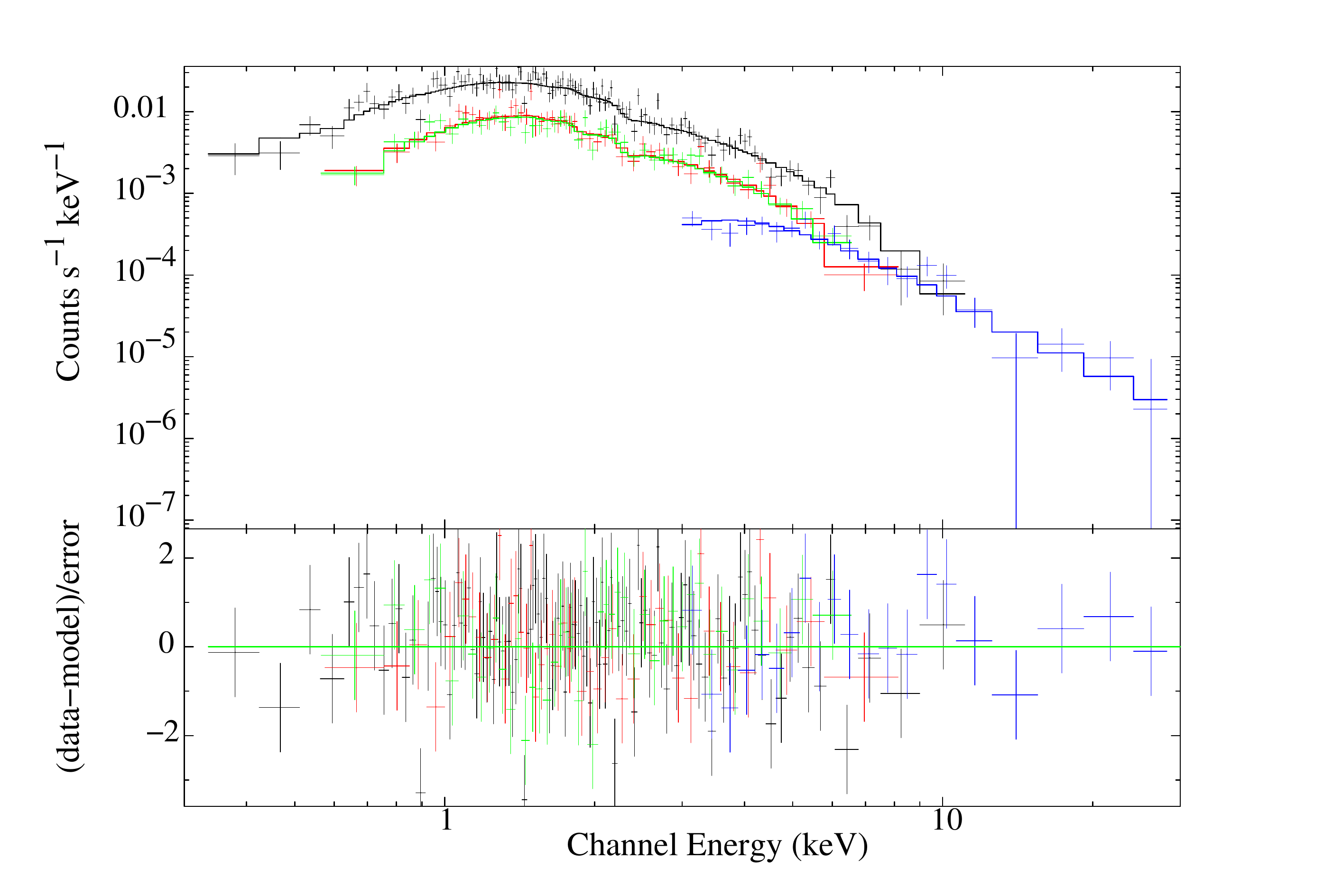}
\caption{{\it XMM-Newton}$+${\it NuSTAR} fitted spectra of X-3 in the 0.3$-$30 keV range. Black, red, green and blue crosses are the EPIC PN, MOS1, MOS2 and {\it NuSTAR} (FPMA$+$FPMB) data, respectively. The spectrum was fitted with diskbb+compTT model. Residual to the diskbb+compTT model is shown in the bottom panel.}
\label{F:xrayfit}
\end{center}
\end{figure*}

We derived the best fitting single-component model parameters for X-3 as given in Table \ref{T:T3}. A two component spectral fitting was also performed with the long exposure data of {\it NuSTAR} (N2) and {\it XMM-Newton} (XM7) in the energy range 0.3$-$30 keV. The best-fitting parameters for two-component model are given in Table \ref{T:T4}. Although, the dates of the two data sets are 9 years apart, the fitting were performed based on the fact that the source does not show significant variability. A similar fitting process were applied by using $N_{H}$ parameters as described above. The obtained spectra are given in Fig. \ref{F:xrayfit}.

In order to search for an underlying pulsar (neutron star), {\it XMM-Newton} EPIC PN data were used to perform timing analyses. The X-ray light curves of X-3 were sampled at 0.1 s and resultant power density spectra (PDS) were calculated. In addition, X-ray light curves were detrended using a 2nd degree polynomial to clean the excessive red noise in lower frequencies in the PDS. The PDS were calculated from single interval or up to 12 spectra were averaged to produce a PDS in the 0.3-10 keV band. 
We have searched for significant peaks in PDS given a continuum red noise level. Significance levels were determined by fitting the power spectra with a two- component model of a Lorentzian and a constant. The significance is calculated as $\sigma$= (P$_{max}$ - P$_{con}$)/P$_{err}$, where P$_{max}$ is the power of the selected peak in the PDS, P$_{con}$ is the continuum value around the peak and P$_{err}$ is the error in the peak value of power \citep{2010MNRAS.407.1895B}. We did not find any peak with a significance large than 1.4$\sigma$ for all PDS we have analysed. Thus, we cannot confirm any periodicity at $\geq$3$\sigma$ confidence level. We note that our time-bin size of 0.1 s sets a lower limit for periods that were searched.

\subsection{Astrometry and {\it HST} Observations}
\label{sssec:optic}

Identification of the optical candidates of X-3 in the NGC 4258 requires accurate source positions. An intercomparison of {\it Chandra}, {\it HST} and {\it SDSS} observations were carried out to obtain improved astrometry. We chose deep {\it Chandra} ACIS observation (ObsID 1618) and {\it HST} observation with Advanced Camera for Surveys (ObsID JB1F89010). The {\it SDSS} (Sloan Digital Sky Survey; \citealp{2015ApJS..219...12A}) {\it r}-band image was also chosen. The {\scshape ciao} tool {\it wavdetect} was used to detect discrete sources on ACIS-S. We selected close, unique, isolated and bright 4 sources both in {\it Chandra} and {\it SDSS} images to be confident in astrometry. These sources seem to be a group but they are not spatially located on specific parts of the CCD. In this case, the vignetting effect is not considered. The matched pairs of objects in the comparisons are presented in Table \ref{T:T5}. The positional uncertainties are given at 90 per cent confidence level for the {\it Chandra}/{\it SDSS} reference sources. The astrometric errors for the {\it Chandra}$-${\it SDSS} are R.A. 0\farcs02 and Dec. 0\farcs15 and {\it SDSS}$-${\it HST} comparisons are R.A. 0\farcs63 and Dec. 0\farcs04. The final corrections used to translate the {\it Chandra} position of X-3 onto the {\it HST} image are 0\farcs65{$\pm$}0\farcs33 in R.A. and 0\farcs18$\pm$0\farcs09 in Dec. These comparisons give the uncertainties that are a quadratic sum of the standard deviations. Then the corrected position of X-3 is determined as R.A. $=$ 12$^{\mathrm{h}}$18$^{\mathrm{m}}$57$^{\mathrm{s}}.90$, Dec. $=$ +47$^{\circ}$16$\arcmin$07$\arcsec.62$ within 90 per cent confidence level of error circle with 0\farcs28 radius. A similar calculation was also used by \cite{2015ApJ...812L..34W}.

We have analysed archival {\it HST} images obtained with the Advanced Camera for Surveys (ACS) given in Table \ref{T:T6}. The PSF photometry was performed using the ACS module in the {\scshape Dolphot}(v2.0, \citealp{2000PASP..112.1383D}). The images were processed by masking all bad pixels using the {\it acsmask} task and the multi-extension *.fits files were split into single-chip images using the {\it Splitgroups} task before performing photometry. Then, the sky background for each chip was calculated with the {\it calcsky} task. We run {\it dolphot} task on both bias and flat-field corrected *.flt and *.drz images. This task was used for photometry on the images by taking the F555W drizzled image as the positional reference in both epochs. The magnitudes in the VEGAmag and Johnson system for the possible optical counterparts are given in Table \ref{T:T7}.

\begin{figure*}
\begin{center}
\includegraphics[width=\textwidth]{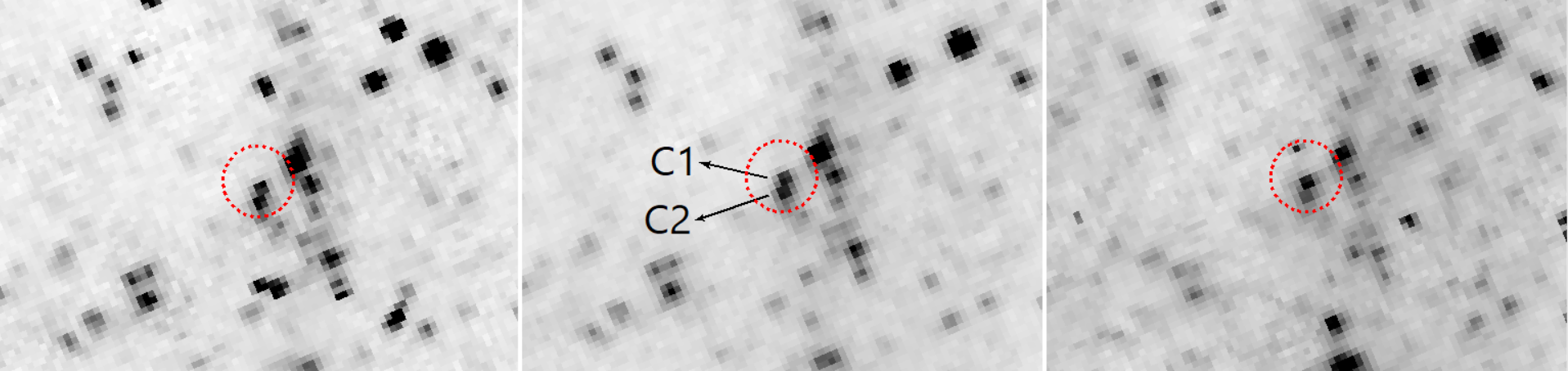}
\caption{The {\it HST}/ACS images of the X-3 in three filters; F814W, F555W and F435W (from left to right). The dashed red circles represent the corrected position of X-3 with an accuracy of 0.\arcsec28 error radius. Two optical candidates (C1 and C2) are shown within the error circle on F555W image. There is a cosmic ray within the error circle on F814W image.}
\label{F:counterpart}
\end{center}
\end{figure*}

After the astrometric correction and photometry, we checked the position of the X-3 to find its optical candidates. Two optical candidates were identified within the error radius of 0\farcs28. The corrected position of X-3 on {\it HST}/ACS images together with the optical candidates are shown in Fig.\ref{F:counterpart}. These candidates are labelled as C1 and C2 according to decreasing Dec. coordinates. 

The Galactic extinction along the direction of NGC 4258 is $E$($B-V$) $\approx 0.014$ mag \citep{2011ApJ...737..103S}. However, there are two extragalactic extinction studies in the literature. The first study from \cite{2013ApJ...779L..20K} derived the $E$($B-V$)$=$0.23$\pm0.03$ using a blue supergiant star in the disk of NGC 4258, 48$\arcsec$ south-east of X-3 (37.8 pc / 1$\arcsec$). The second study from \cite{2006ApJ...652.1133M} gave 281 Cepheid stars with $E$($B-V$) values in the range of 0.01 $-$ 0.44 in the NGC 4258. According to the second study, we obtained the extragalactic extinction as $E$($B-V$)$\approx$0.17 by selecting about thirty Cepheids close to X-3 region ($\approx$160$\arcsec$ north-west). The standard deviation of the extinction from the Cepheids is calculated as 0.08 mag. These two extragalactic extinction values ($0.17$ and $0.23$) were used to determine the spectral type of the optical candidates. Then, both values yielded compatible results. Therefore, the average value $E$($B-V$)$=$0.20 was adopted as extinction value throughout the paper. The extinction corrected magnitudes and color values of the optical candidates are given in Table \ref{T:T7}. The average standard deviation (0.08) does not affect the counterparts' features significantly.

\begin{figure*}
\begin{center}
\includegraphics[angle=0,scale=0.30]{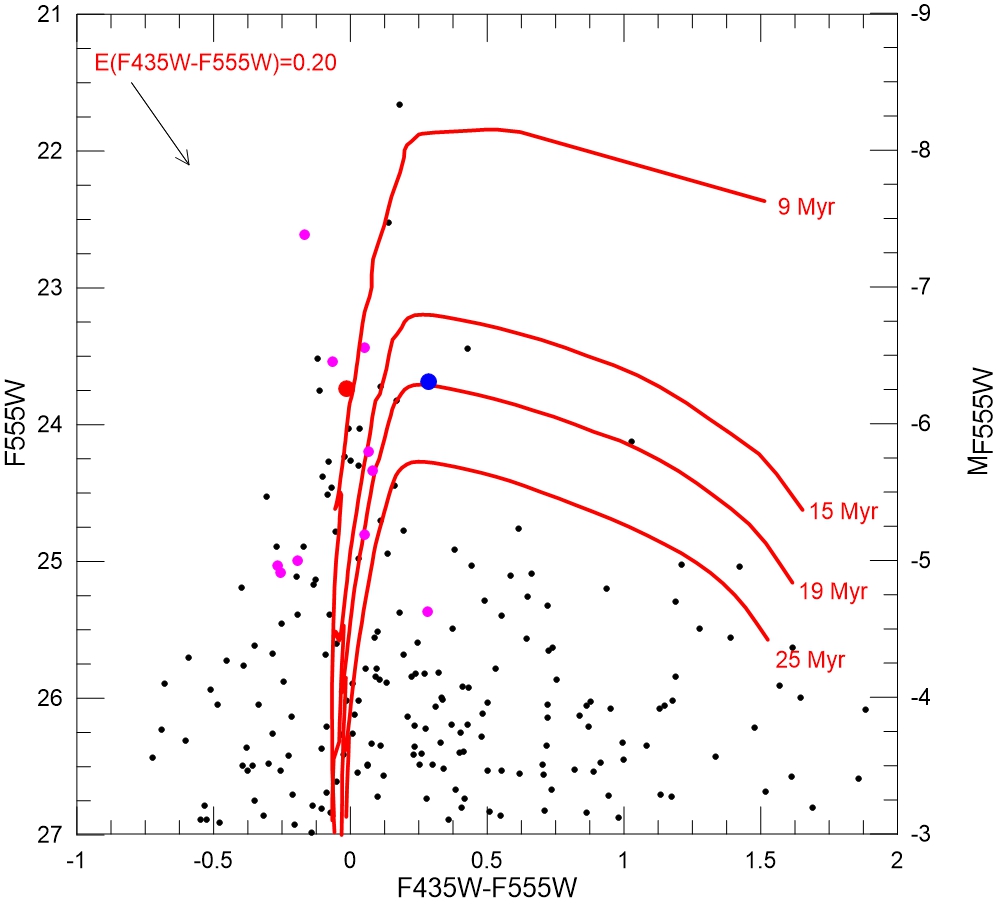}
\includegraphics[angle=0,scale=0.30]{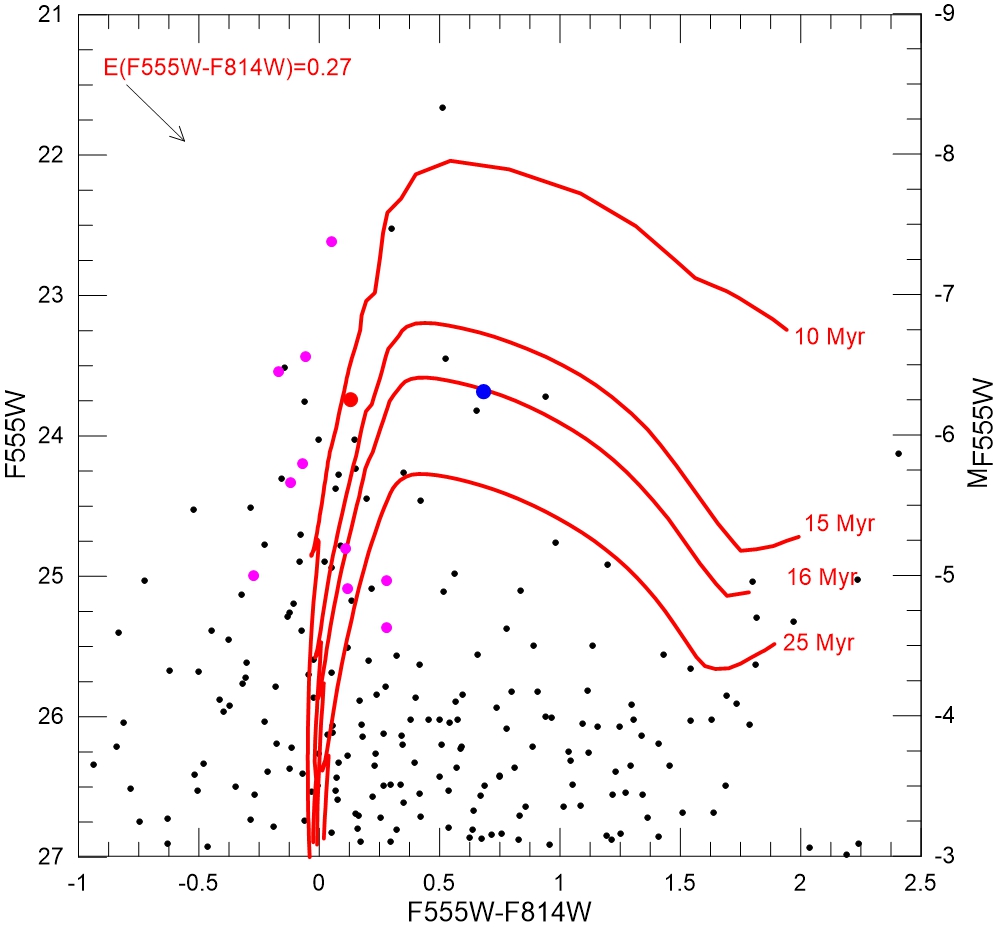}
\caption{The {\it HST}/ACS color$-$magnitude diagrams (CMDs) for optical candidates and field stars around the X-3. Padova isochrones of different ages are overplotted. The blue and red dots represent optical candidates C1 and C2, respectively. The black and magenta dots represent field stars within the 25 arcsec$^{2}$ square region around the X-3 and nearby stars, respectively. These isochrones have been corrected for extinction of A$_{V}$ $=$ 0.62 mag and the black arrows shows the reddening line.} 
\label{F:age}
\end{center}
\end{figure*}

\begin{figure*}
\begin{center}
\includegraphics[angle=0,scale=0.30]{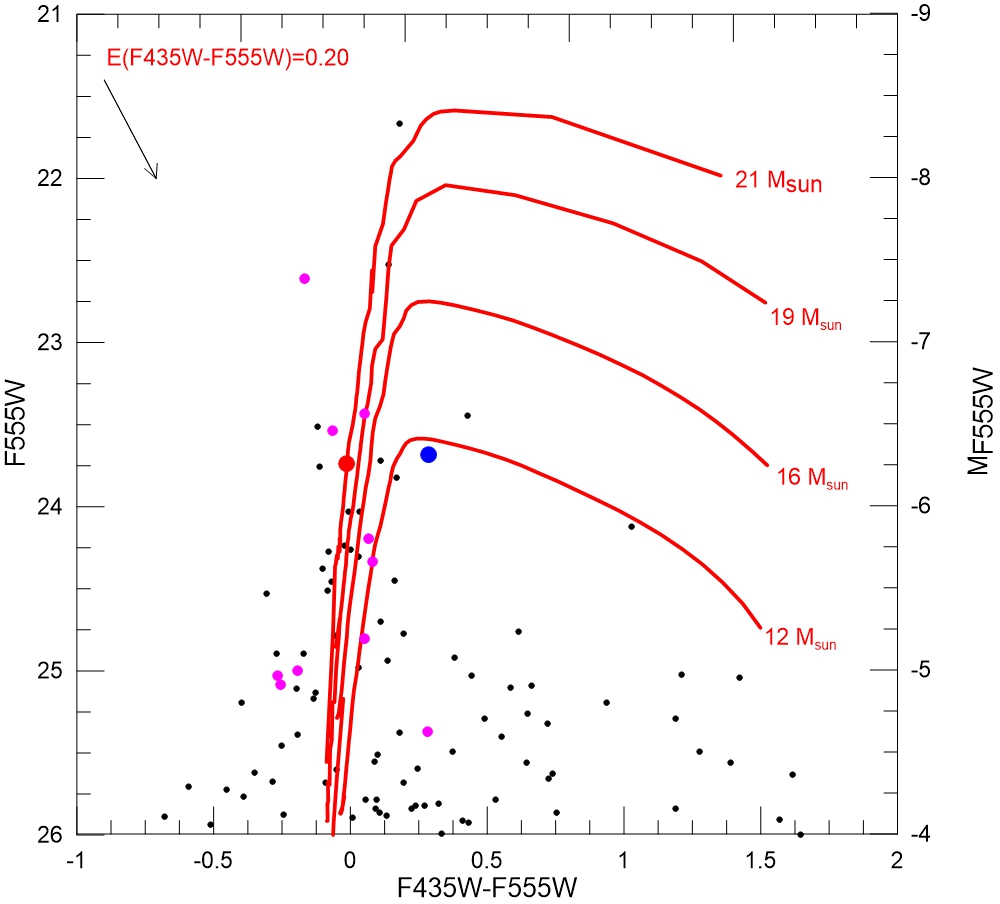}
\includegraphics[angle=0,scale=0.30]{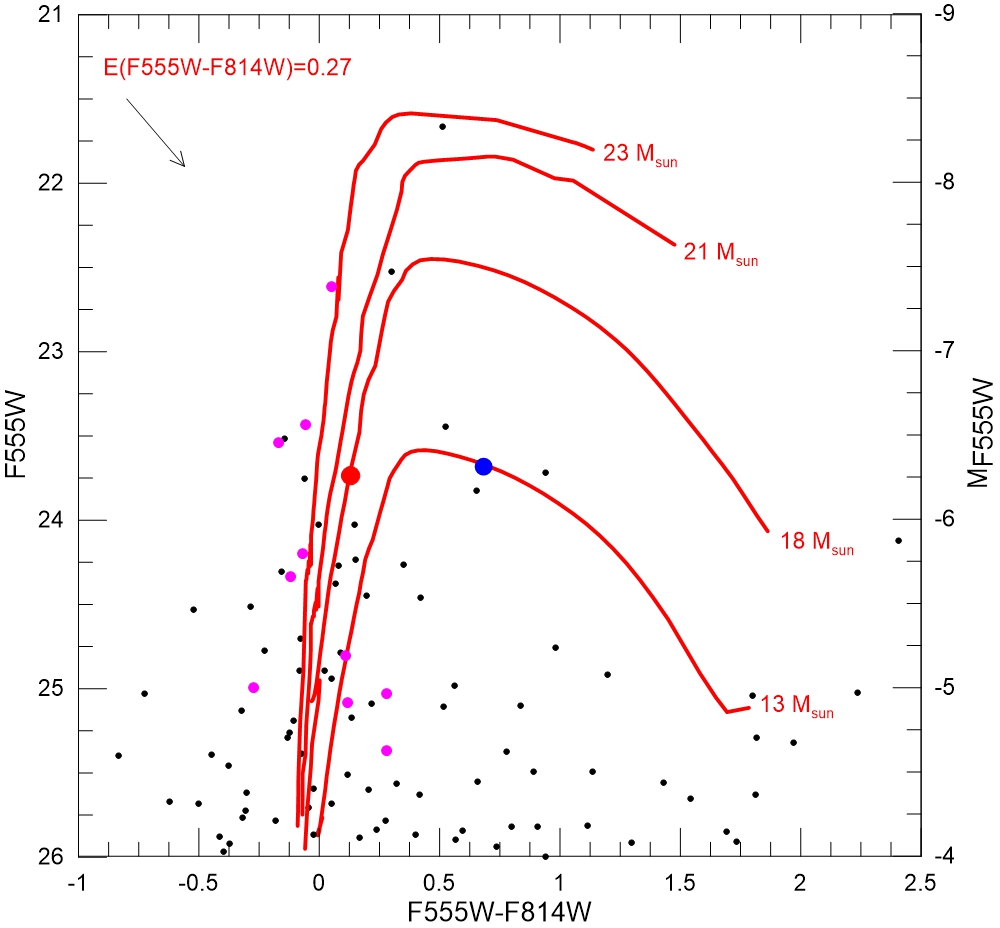}
\caption{The {\it HST}/ACS CMDs for optical candidates and field star around the X-3. Padova isochrones of different masses are overplotted. Definitions are the same as in Fig. \ref{F:age}.}
\label{F:mass}
\end{center}
\end{figure*}

In order to estimate the age and mass of the optical candidates, the color-magnitude diagrams (CMDs) of X-3 and its environment were obtained. Two CMDs as F555W versus F435W$-$F555W and F814W versus F555W$-$F814W were derived for optical candidates and the field stars. These stars within the 25 arcsec$^{2}$ square region around X-3 is shown in Fig. \ref{F:ngc4258}. The metallicity of NGC 4258 was used as Z$=$0.011 from \cite{2013ApJ...779L..20K} to obtain for the corresponding PARSEC isochrones. In Fig. \ref{F:age} and Fig. \ref{F:mass}, the age and mass isochrones have been overplotted on the CMDs. The distance modulus was calculated as 29.4 magnitude using the adopted distance 7.7 Mpc.

\begin{figure}
\begin{center}
\includegraphics[width=\columnwidth]{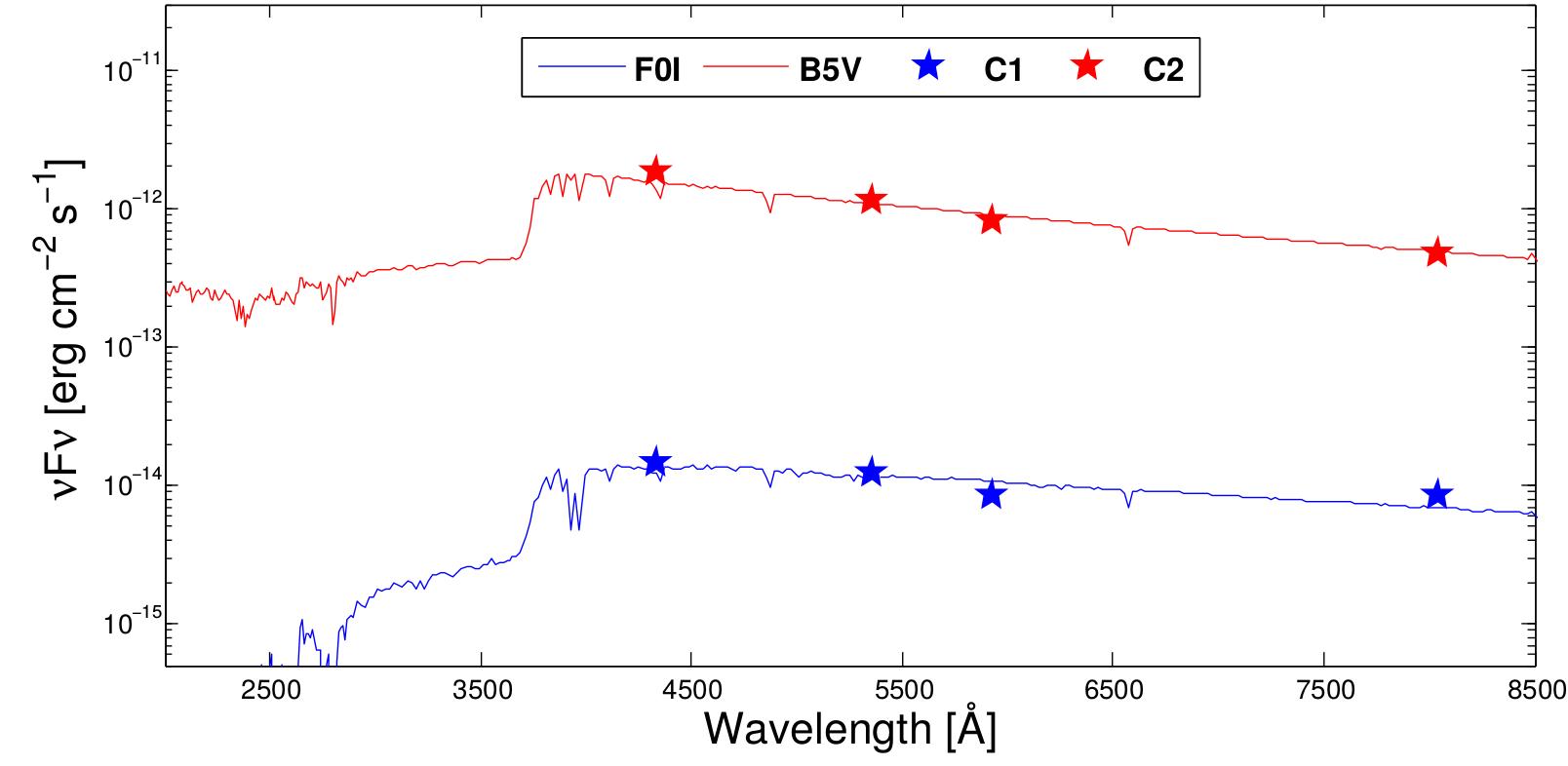}
\caption{The SEDs of optical candidates C1 and C2. The blue and red lines represent the synthetic spectra derived with metallicity of Z $=$ 0.011 and extinction of A$_{V}$ $=$ 0.62 mag for F0I and B5V, respectively. The blue and red circles represent flux values of candidates for C1 and C2, respectively. There is a systematic error less than 3 per cent. The red line was shifted upward by factor of a hundred for clarity.}
\label{F:sp}
\end{center}
\end{figure}

The spectral types of C1 and C2 were estimated using Spectral Energy Distribution (SED) templates with the {\scshape pysynphot }\footnote{https://pysynphot.readthedocs.io/en/latest/} using the CK04 standard stellar spectra \citep{2004A&A...419..725C}. The SEDs for C1 and C2 are constructed for all flux values which are derived from {\it HST}/ACS magnitudes in Table \ref{T:T7}. Synthetic spectra are normalized to $m_{V}$ = 0 mag. The reduced $\chi^{2}$ of the best fits are 3.93 and 2.26, respectively. The resultant SEDs of the C1 and C2 were found using the best-fit models given in Fig. \ref{F:sp}.

\section{Discussion and Conclusions}
\label{sec:discussion}

In this study, the archival X-ray data of the X-3 in the nearby galaxy NGC 4258 were analysed. There are 10 X-ray observations of the source obtained by {\it XMM-Newton}, {\it Chandra} and {\it NuSTAR} observatories covering $\approx 15$ years. Also the optical properties of the source were studied with the {\it HST} observations. Two optical candidates were found within 0\farcs28 error radius and they were further examined.

In the spectral analyses of the {\it XMM-Newton} and {\it Chandra} data, resultant spectral fits obtained show a wide range of $\chi_{\nu}^{2}$ values going from 0.7 (which means that the data could be under-sampled) to 1.8 (which means that the fit is not a good one). If we consider the application of simple phenomenological models for the obtained spectra of the source, we note that the diskbb model fits better than the pl model on a 3$\sigma$ confidence level for XM5 and XM7 datasets (according to F-test). On the other hand, the spectra of XM1 and XM2 fit with the pl model better at 2-3$\sigma$. However, it is not possible to distinguish these two models for the remaining datasets; XM3, XM4, XM6 and Ch1.

Using the results from \citet{2009MNRAS.397.1836G}, we elaborate on some of the spectral characteristics of X-3. The spectra of XM1 and XM2 are better represented by the pl model with photon index $\Gamma$ $\sim$ (2 $-$ 2.2). These $\Gamma$ values correspond to hard states defined for Galactic BHBs. Hard state with low luminosity is seen at sub-Eddington mass accretion rate. On the other hand, the diskbb model yields acceptable fits for XM5 and XM7 with the temperature range of $kT_{in}$ $\sim$ (1.09 $-$ 1.33) keV. These $kT_{in}$ values are compatible with those of Galactic BHBs at a high mass accretion rate during the thermal state \citep{2006ARA&A..44...49R}. Generally in Galactic BHBs, luminosities are usually higher in the thermal state than the hard state. However, we might interpret that X-3 exhibits the opposite behavior since the source has a high $L_{X}$ when it is in hard state and a low $L_{X}$ when it is in a thermal state. There are some ULXs that do show similar behavior: NGC 1313 X-2, \citet{2006ApJ...650L..75F}; IC 342 X-1, \citet{2014MNRAS.444..642M}; NGC 4736 X-2, \citet{2014Ap&SS.352..123A}. As discussed in several studies, when the data quality is low and exposure is short, one-component models should be taken into account statistically. However, these models do not provide physically sufficient evidence to interpret the data \citep{2009MNRAS.397.1836G, 2013MNRAS.435.1758S,2017ARA&A..55..303K}. Therefore, we combined two continuum models (pl and diskbb) and fitted all data sets accordingly to examine the spectral characteristics and classification according to the prescription of \citet{2013MNRAS.435.1758S}. In their work, an empirical scheme is used to classify ULXs into three classes due to their spectral morphology, which are a broadened disk, a two-component hard ultraluminous and a two-component soft ultraluminous classes.
When, we apply the two-component model to available data, we were only able to obtain physically meaningful parameters for XM7. The spectrum of XM7 is adequately fitted with the doubly absorbed (diskbb + pl) spectral model ($\Gamma$=0.87 and $kT_{in}$ $\sim$ 1.21 keV) with a $\chi_{\nu}^{2}$ $\sim$ 1.01 as given in Table \ref{T:T2}. Due to the very flat photon index $\Gamma$, it is fixed to values between 1.7 and 2 (no acceptable fit outside this range) however, $kT_{in}$ and $\chi_{\nu}^{2}$ values were not changed significantly. The best-fit temperature parameter ($kT_{in}$ = 1.21 keV) is > 0.5 keV, based on the chart in the form of the decision tree in the Figure 2 of \citet{2013MNRAS.435.1758S}, the calculated $F_{pl}$/$F_{disk}$ ratio is found to be 0.29 ($\Gamma$ was fixed 1.7 while calculating F$_{pl}$) which indicates that the X-3 spectrum can be classified as a “broadened disk”. They defined that broadened disk class ULX population has $L_{X}$ < 3 $\times$ $10^{39}$ erg $s^{-1}$ and this is consistent with a population of stellar mass black hole (M < 20 M$_{\sun}$) accreting at just above the Eddington limit. The spectra of this class are thought to be dominated by the accretion-disk, but due to the high accretion rate the disk structure modified from the standard thin disk. 

Pulsating ULXs (PULXs) are known to show luminosity variabilities at least a factor of 100 \citep{2019arXiv190604791R, 2017Sci...355..817I, 2017MNRAS.466L..48I, 2016ApJ...831L..14F}. However, the source X-3 exhibits a variability usually less than factor of 3 throughout the observations spanning 15 years. This may be another clue for a black hole as a compact accretor rather than a neutron star. Then, by using the diskbb model's normalization parameter N = (R$_{in}$ / D$_{10}$)$^{2}$ $\times$ cos$\theta$ (defined by \citealp{2000ApJ...535..632M}), the mass of the compact object in the system can also be estimated. In the formula, $R_{in}$ is the inner disk radius in km, $D_{10}$ is the distance to the source in units of 10 kpc, and $\theta$ is the inclination angle of the disk.
We calculated the mass using the best-fit normalization parameter of diskbb model ($N_{\mathrm{disk}} = 5.5\pm0.2 \times 10^{-3}$) derived for the longest exposure data, XM7. The inner disk radius was found to be as $R_{\mathrm{in}}$ $=$ 95 km (using the equation $R_{\mathrm{in}}= $ $\xi \kappa^{2} r_{\mathrm{in}}$ where the correction factor $\xi=0.41$, spectral hardening factor $\kappa=1.7$, and $r_{\mathrm{in}}$ is the apparent inner disk radius; see (\citealp{1995ApJ...445..780S, 1998PASJ...50..667K}). Assuming a moderate disk inclination of $60^{\circ}$, we can calculate the mass of the compact object as $\sim$10 M$_{\sun}$ indicating a stellar mass black hole. 

We also used the data from {\it NuSTAR} observations (which has the energy range of 3 $-$ 79 keV) for further understanding of the X-3 system. Although the {\it NuSTAR} observations are not simultaneous with {\it XMM-Newton} in time, due to low variability of X-3 system, the joint analysis of both data is performed to examine the emission of the source in the high energies ($>$10 keV).
The spectral results using one-component model fits are summarized in Table \ref{T:T3}. The models provide slightly better fits with a diskpbb or a cutoffpl model with a luminosity $L_{x}$ $\sim$ 2.7$\times$ $10^{39}$ erg $s^{-1}$. We also investigated whether the spectral fits are improved over that of a one-component model fits by adding a second component in modelling. We found that derived E$_{cut}$ values for one and two component models are in accordance with the findings in the literature \citep{2015ApJ...799..121R, 2016MNRAS.463..756K, 2017ApJ...839...46S, 2019A&A...621A.118K}. The results are given in Table \ref{T:T4}. This joint analysis was constructed to test the differences in spectral characteristics of ULXs and PULXs as in the study by \cite{2018ApJ...869..111W}. Diskbb/diskpbb model was used along with either a compTT model for a black hole interpretation where a seed-in temperature from the disk is connected with the Comptonized plasma in/above the disk or a cutoffpl model was assumed for a neutron star accretor interpretation where the emission is from the accretion column.
The temperature parameter of compTT is fixed to 50 keV while applying the diskpbb+compTT model to the spectra. This temperature is similar to that seen in Galactic BHBs while they are in classical accretion states \citep{2009MNRAS.397.1836G, 2013ApJ...778..163B}. The p parameters were found to be consistent with the standard disk model (p $\sim$ 0.75). In this case, models containing the disk model (with one or two components) may be more favorable.
Nevertheless, these models did not yield any distinguishable fits considering reduced $\chi^{2}$ values and null-hypothesis probabilities. In addition, F-test probability values ($\sim$ $10^{-5}$) did not show any significant improvement of the fits over the one-component models.
As a result, {\it NuSTAR} data jointly analysed with the {\it XMM-Newton} data did not help to elaborate further our ULX model for X-3 which predicts a black hole as a compact accretor.

Using the usual CMD techniques, the derived mean age values of the optical candidates C1 and C2 are $\approx$ 18 Myr and $\approx$ 10 Myr, respectively as described in section \ref{sssec:optic}. Assuming the optical emission is dominated by the donor star, spectral types of the possible optical counterparts are found to be A3 $-$ F1 giant for C1, and B3 $-$ B6 main sequence star for C2. Similarly, the SEDs from stellar templates (CK04) of C1 and C2 candidates match with spectral types as F0I and B5V, respectively. 
Their masses can also be constrained using the mass isochrones as $\approx$ 13 M$_{\sun}$ and 20 M$_{\sun}$, for C1 and C2, respectively. Both optical candidates of X-3 are young blue stars with $(B-V)_{0}$ values in the range of -0.13$-$0.16.

There is a distinct arm in the region of X-3 where new star formations seem quite dense as seen from Fig \ref{F:ngc4258}. We also investigate the environment of X-3 within 25 arcsec$^{2}$ region.
Although, there was no obvious star cluster near the ULX source, the selected 12 stars with magenta color represent the bright stars very close to X-3 shown in Fig. \ref{F:age} and Fig. \ref{F:mass}.
These bright stars in the CMDs have young ages of < 40 Myr and their (B-V) colors are between -0.4 and 0.4 with m$_{V}$ $\approx$ 22.5$-$25.5 mag. The ages and color values of the candidates of X-3 and the stars in the selected region seem compatible with each other.

In order to differentiate the optical candidates from background AGNs, we obtained their X-ray to optical flux ratios. We used XM7 and {\it HST}/ACS F555W (ObsID JB1F89010) data sets for these calculation since simultaneous observations are not available for these wavelengths. $F_{\mathrm{X}}/F_{\mathrm{opt}}$ values for C1 and C2 were found as 390 and 360, respectively. On the other hand, ratios for active galactic nuclei are in the range of 0.1 $\leq$ $(F_{x} / F_{opt})_{AGN}$ $\leq$ 10 \citep{2010MNRAS.401.2531A}. These ratio values of optical candidates of X-3 are significantly higher than those of AGNs but are compatible with known ULXs: 260 < $F_{X}$/$F_{opt}$ < 4200 \citep{2016MNRAS.455L..91A, 2016ApJ...828..105A}. 

We believe the ULX source, X-3 in NGC 4258, will make use of broadband X-ray spectra obtained at higher sensitivity and observation sampling to study the source spectra and its possible spectral variations. Further joint longer observations using {\it XMM-Newton} and {\it NuSTAR} may allow us to constrain the characteristics of the X-ray emission components and perform better timing analyses. The ongoing missions like {\it eROSITA}/SRG and the future missions like {\it Athena} can resolve the complexities we have outlined in our work.

\section*{Acknowledgements}

We thank the anonymous referee for providing very helpful comments. We also thank M.E. Özel for his valuable contributions and suggestions. This research was supported by the Scientific and Technological Research Council of Turkey (TÜBİTAK) through project number 117F115. This research was also supported by the Çukurova University Research Fund through project number FBA-2019-11803.

\section*{Data Availability}
The scientific results reported in this article are based on archival observations made by the Chandra X-ray Observatory,  as well as archival observations by XMM-Newton, an ESA science mission with instruments and contributions directly funded by ESA Member States and NASA, archival observations by the NuSTAR mission, a project led by the California Institute of Technology, managed by JPL, and funded by NASA. This work has also made use of observations made with the NASA/ESA Hubble Space Telescope, and obtained from the data archive at the Space Telescope Science Institute. STScI is operated by the Association of Universities for Research in Astronomy, Inc. under NASA contract NAS 5-26555."

\bibliographystyle{mnras}
\bibliography{ngc4258} 

\begin{table*}
\centering
\caption{The log of {\it XMM-Newton}, {\it Chandra} and {\it NuSTAR} observations.}
\begin{tabular}{cccccc}
\hline\hline
& Label & ObsID & Date & Exp. \\
& & & & (ks)\\
\hline
{\it XMM-Newton} & XM1 & 0110920101 & 2000-12-08 & 23.31 \\
		 & XM2 & 0059140101 & 2001-05-06 & 12.70 \\
		 & XM3 & 0059140201 & 2001-06-17 & 13.16 \\
		 & XM4 & 0059140401 & 2001-12-17 & 15.01 \\
		 & XM5 & 0059140901 & 2002-05-23 & 16.51 \\
		 & XM6 & 0203270201 & 2004-06-01 & 48.91 \\
		 & XM7 & 0400560301 & 2006-11-18 & 64.52 \\
\hline
{\it Chandra} & Ch1 & 700234  & 2001-05-28 & 20.94 \\
\hline
{\it NuSTAR} & N1 & 60101046002 & 2015-11-16 & 54.78 \\ 	 
			 & N2	 & 60101046004 & 2016-01-10 & 103.6 \\
\hline
\end{tabular}
\label{T:T1}
\end{table*}

\begin{table*}
\centering
\begin{minipage}[b]{0.9\linewidth}
\caption{Best-fitting spectral parameters for X-3 from the {\it XMM-Newton} and {\it Chandra} observations. The errors are at 90 per cent for each parameter.}
\begin{tabular}{cccccccccccc}
\hline\hline
No. & N$_{H}$ & $\Gamma$ \footnote[1]{photon index of the pl component}& $\Gamma$* & kT$_{\mathrm{in}}$ \footnote[2]{inner disk temperature} & kT$_{\mathrm{in}}$* &$\chi_{\nu}^{2}$ (dof) \footnote[3]{reduced $\chi_{\nu}^{2}$} & $\chi_{\nu}^{2}$* (dof)& L$_{\mathrm{X}}$ \footnote[4]{Unabsorbed luminosity values were calculated using a distance of 7.7 Mpc at 0.3$-$10 keV energy range.} & L$_{\mathrm{X}}$* \\
- & ($10^{22}$ cm$^{-2}$) & - & & (keV) & (keV) & - &  & (10$^{39}$ erg s$^{-1}$) & (10$^{39}$ erg s$^{-1}$) & \\
\hline
\multicolumn{6}{c}{tbabs $\times$ pl}\\
XM1 & $0.39_{-0.04}^{+0.04}$ & $2.15_{-0.14}^{+0.14}$ & $2.43_{-0.16}^{+0.16}$ & - & - &1.24 (35) & 1.29 (38) & $1.83_{-0.17}^{+0.17}$ & $2.31_{-0.17}^{+0.17}$ \\
XM2 & $0.55_{-0.06}^{+0.07}$ & $2.27_{-0.15}^{+0.15}$ & $2.23_{-0.15}^{+0.16}$ & - & - &1.36 (24) & 1.39 (27) & $4.19_{-0.47}^{+0.47}$ & $3.52_{-0.47}^{+0.47}$ \\
Ch1 & $0.60_{-0.05}^{+0.05}$ & $2.00_{-0.12}^{+0.12}$ & $1.86_{-0.11}^{+0.12}$ & - & - &0.72 (34) & 0.74 (35) & $5.38_{-0.31}^{+0.32}$ & $4.93_{-0.31}^{+0.32}$ \\ 
XM3 & $0.40_{-0.05}^{+0.06}$ & $2.20_{-0.15}^{+0.15}$ & $2.39_{-0.16}^{+0.17}$ & - & - &0.79 (27) & 0.91 (30) & $2.38_{-0.26}^{+0.25}$ & $3.07_{-0.26}^{+0.25}$ \\
XM4 & $0.65_{-0.07}^{+0.09}$ & $2.21_{-0.17}^{+0.18}$ & $1.99_{-0.16}^{+0.17}$ & - & - &0.92 (73) & 0.92 (76) & $5.20_{-0.63}^{+0.62}$ & $4.17_{-0.63}^{+0.62}$ \\
XM5 & $0.61_{-0.06}^{+0.07}$ & $2.04_{-0.11}^{+0.12}$ & $1.93_{-0.11}^{+0.11}$ & - & - &1.62 (47) & 1.55 (50) & $3.53_{-0.30}^{+0.30}$ & $3.32_{-0.30}^{+0.30}$ \\
XM6 & $0.36_{-0.04}^{+0.04}$ & $2.22_{-0.15}^{+0.16}$ & $2.56_{-0.18}^{+0.19}$ & - & - &0.71 (57) & 0.84 (60) & $1.87_{-0.19}^{+0.19}$ & $2.73_{-0.19}^{+0.19}$ \\
XM7 & $0.53_{-0.02}^{+0.02}$ & $2.15_{-0.05}^{+0.05}$ & $2.13_{-0.05}^{+0.05}$ & - & - & 1.27 (229) & 1.27 (232) & $3.92_{-0.14}^{+0.14}$ & $3.90_{-0.14}^{+0.14}$ \\
\hline
\multicolumn{6}{c}{tbabs $\times$ diskbb}\\
XM1 & $0.16_{-0.03}^{+0.04}$ & - & - & $1.12_{-0.10}^{+0.12}$ & $1.00_{-0.11}^{+0.14}$ & 1.81 (35) & 1.78 (38) & $0.97_{-0.09}^{+0.09}$ & $1.01_{-0.09}^{+0.09}$\\
XM2 & $0.22_{-0.05}^{+0.07}$ & - & - & $1.16_{-0.12}^{+0.14}$ & $1.15_{-0.13}^{+0.15}$ & 1.52 (24) & 1.53 (27) & $2.02_{-0.23}^{+0.23}$ & $1.80_{-0.23}^{+0.23}$ \\
Ch1 & $0.32_{-0.04}^{+0.05}$ & - & - & $1.34_{-0.10}^{+0.11}$ & $1.50_{-0.15}^{+0.18}$ & 0.70 (34) & 0.79 (35) & $3.14_{-0.18}^{+0.18}$ & $3.04_{-0.18}^{+0.18}$ \\
XM3 & $0.10_{-0.05}^{+0.06}$ & - & - & $1.18_{-0.11}^{+0.13}$ & $1.03_{-0.12}^{+0.14}$ & 0.94 (27) & 1.18 (30) & $1.22_{-0.13}^{+0.13}$ & $1.42_{-0.13}^{+0.13}$ \\
XM4 & $0.33_{-0.06}^{+0.08}$ & - & - & $1.16_{-0.13}^{+0.15}$ & $1.32_{-0.18}^{+0.22}$ & 0.91 (73) & 0.92 (76) & $2.64_{-0.32}^{+0.31}$ & $2.39_{-0.32}^{+0.31}$ \\
XM5 & $0.29_{-0.05}^{+0.06}$ & - & - & $1.33_{-0.10}^{+0.12}$ & $1.42_{-0.13}^{+0.15}$ & 1.26 (47) & 1.22 (50) & $2.04_{-0.17}^{+0.17}$ & $2.02_{-0.17}^{+0.17}$ \\
XM6 & $0.13_{-0.03}^{+0.04}$ & - & - & $1.09_{-0.10}^{+0.12}$ & $0.94_{-0.12}^{+0.14}$ & 0.78 (57) & 0.92 (60) & $0.98_{-0.10}^{+0.10}$ & $1.13_{-0.10}^{+0.10}$ \\
XM7 & $0.25_{-0.02}^{+0.02}$ & - & - & $1.29_{-0.04}^{+0.05}$ & $1.33_{-0.05}^{+0.05}$ & 1.02 (229)& 1.03 (232)& $2.24_{-0.08}^{+0.08}$ & $2.23_{-0.08}^{+0.08}$ \\
\hline
\multicolumn{6}{c}{tbabs $\times$ (po+diskbb)}\\
XM7** & $0.25_{-0.02}^{+0.02}$ & $0.87_{-0.22}^{+0.37}$ &-- & $1.21_{-0.01}^{+0.01}$ & -- & 1.02 (228) & -- & $2.23_{-0.01}^{+0.01}$ & -- & \\
\hline
\end{tabular}
\label{T:T2}
\\ * These values were calculated using a fixed N$_{H}$ value as 0.51 $\times$ 10$^{22}$ cm$^{-2}$ for the pl and 0.22 $\times$ 10$^{22}$ cm$^{-2}$ for the diskbb models.
\\ ** Normalization parameters are 1.76 $\times$ 10$^{-6}$ photon/keV/cm$^{2}$/s at 1 keV for pl model and 6.58 $\times$ 10$^{-3}$ [(r$_{in}$ km$^{-1}$)/(D/10 kpc)]$^{2}$ $\times cosi$ for diskbb model.  
\end{minipage}
\end{table*}

\begin{table*}
\centering
\begin{minipage}[b]{0.9\linewidth}
\caption{Spectral parameters obtained with one-component model fits for X-3 in the {\it XMM-Newton}+{\it NuSTAR} (N2)}
\begin{tabular}{c c c c c c c c l }
\hline
 model   & N$_{H}$              &   $\Gamma$ & kT$_{\mathrm{in}}$ /kT$_{\mathrm{e}}$ \footnote[1]{electron temperature of the corona}  & E$_{cut}$ \footnote[2]{e-folding energy of the cutoffpl model} &   p \footnote[3]{exponent of the radial dependence of the disk temperature}  & $\tau$ \footnote[4]{optical depth of the corona} & $\chi^{2}$/dof &  L$_{\mathrm{X}}$ (10$^{39}$)          \\
         & $(10^{22})$ cm$^{-2}$ &            &  keV &        keV        &        &   &  &  erg s$^{-1}$   \\
\hline
pl       & $0.58_{-0.02}^{+0.02}$ & $2.27_{-0.03}^{+0.04}$ & - &  -    &   -   & -  & 300.18/232 (1.29) & 4.67  \\
diskbb   & $0.25_{-0.01}^{+0.02}$ & - &$1.31_{-0.01}^{+0.01}$ &  -   &   -   & -  & 260.10/232 (1.12)  &  2.29  \\
diskpbb  & $0.37_{-0.02}^{+0.02}$ & -&$1.60_{-0.01}^{+0.01}$ & -& $0.58_{-0.003}^{+0.004}$ &    -  &  251.75/231 (1.09)  & 2.78  \\
compTT   & $0.57_{-0.02}^{+0.02}$ & &$53.44_{-0.89}^{+0.88}$ &  -    & -& $0.34_{-0.01}^{+0.01}$ & 296.58/230 (1.29)  & 3.77 \\
cutoffpl & $0.33_{-0.02}^{+0.02}$ & $0.84_{-0.03}^{+0.04}$ &-& $2.49_{-0.08}^{+0.08}$  & - & - & 248.26/231 (1.07) &  2.62 \\
\hline
\label{T:T3}
\end{tabular}
\end{minipage}
\end{table*} 

\begin{table*}
\centering
\begin{minipage}[b]{0.9\linewidth}
\caption{{Spectral parameters obtained with two-component model fits  for X-3 in the \it XMM-Newton} (XM7)+{\it NuSTAR} (N2) }
\begin{tabular}{cccccccc}
\hline\hline
Parameter & unit & diskbb+compTT & diskbb+cutoffpl & diskpbb+compTT & diskpbb+cutoffpl \\
\hline
$N_{H}$ & $10^{22}$ cm$^{-2}$ & $0.31_{-0.02}^{+0.02}$ & $0.28_{-0.02}^{+0.02}$ & $0.29_{-0.02}^{+0.02}$ & $0.28_{-0.02}^{+0.02}$ \\
k$T_{in}$ & keV & $1.15_{-0.01}^{+0.01}$ & $1.12_{-0.01}^{+0.01}$& $1.15_{-0.01}^{+0.01}$ & $1.12_{-0.01}^{+0.01}$ \\
p & & -- & -- & $0.75_{-0.01}^{+0.01}$ & $0.75_{-0.01}^{+0.01}$ \\
N\footnote{Normalization parameters of diskbb and diskpbb models. N=[(r$_{in}$ km$^{-1}$)/(D/10 kpc)]$^{2}$ $\times cosi$.} & $10^{-3}$ & $6.93_{-0.32}^{+0.32}$ & $8.09_{-0.33}^{+0.34}$ & $60.7_{-3.22}^{+3.20}$ &$ 8.37_{-0.35}^{+0.35}$ \\
k$T_{e}$ & keV & $45.00_{-3.46}^{+3.45}$ & -- & 50 & \\
$\tau$ & & $1.17_{-0.16}^{+0.19}$ &-- & $0.95_{-0.22}^{+0.31}$ & -- \\
$\Gamma$ & & -- & $0.98_{-0.09}^{+0.11}$ & -- & $1.04_{-0.09}^{+0.11}$ \\
$E_{cut}$ & keV & -- & $15.39_{-4.56}^{+7.92}$ & -- & $16.14_{-4.88}^{+8.80}$ \\
N\footnote{Normalization parameters of compTT and cutoffpl model in units of photon/keV/cm$^{2}$/s at 1 keV.} &$10^{-6}$ & $2.05_{-0.29}^{+0.29}$ &$ 6.11_{-1.12}^{+1.15}$ & $0.8_{-0.1}^{+0.1}$ & $6.87_{-1.21}^{+1.25}$ \\
$\chi^{2}$/dof & & 1.04 (236.39/228) & 1.03 (235.72/229) & 1.03 (236.29/228) & 1.03 (235.76/228) \\
Null P & & 0.34 & 0.45 & 0.39 & 0.27 \\
$L_{\mathrm{X}}$\footnote{Luminosity values were calculated at 0.3$-$30 keV energy range.} & $10^{39}$ erg s$^{-1}$ & $1.69_{-0.02}^{+0.01}$ & $2.58_{-0.01}^{+0.01}$ & $2.33_{-0.01}^{+0.01}$ & $2.78_{-0.02}^{+0.01}$ \\
\hline
\end{tabular}
\label{T:T4}
\end{minipage}
\end{table*}

\begin{table*}
\centering
\begin{minipage}[b]{0.9\linewidth}
\caption{Coordinates and their uncertainties of the X-Ray/Optical reference sources and ULXs.}
\begin{tabular}{ccccccccccc}
\hline\hline
\multicolumn{6}{c}{{\it Chandra} ACIS X-ray sources (ObsID 1618) identified in {\it SDSS} observation (band {\it r})}\\
\hline
{\it Chandra} R.A.&{\it Chandra} Dec.&{\it SDSS} R.A.&{\it SDSS} Dec.& Position Uncertainty (\arcsec)\footnote{The uncertainties are given at 90 per cent confidence level of the {\it Chandra}/{\it SDSS} reference sources.} & Counts \footnote{The background subtracted counts were calculated in the 0.3-10 keV using  {\scshape xspec}.}\\
\hline
12:18:49.488&+47:16:46.56&12:18:49.478&+47:16:46.47&0.166 &  140\\
12:18:59.335&+47:18:20.59&12:18:59.388&+47:18:20.45&0.807 &  11\\
12:18:56.165&+47:18:58.43&12:18:56.119&+47:18:58.14&0.736 &  13\\
12:18:57.506&+47:18:14.47&12:18:57.504&+47:18:14.38&0.095 &  2900\\
\hline
\multicolumn{6}{c}{{\it SDSS} sources ({\it r} band) identified in {\it HST} ACS/WFC/F555W observation of JB1F89010}\\
\hline
{\it SDSS} R.A.&{\it SDSS} Dec.&{\it HST} R.A.&{\it HST} Dec.&\\
\hline
12:18:55.032&+47:15:53.57&12:18:55.075&+47:15:53.63&0.651 & --\\
12:18:57.528&+47:15:30.28&12:18:57.578&+47:15:30.46&0.741 & --\\
12:18:55.838&+47:15:34.23&12:18:55.874&+47:15:34.34&0.551 & --\\
12:18:54.110&+47:15:37.83&12:18:54.149&+47:15:37.92&0.579 & --\\
12:18:53.750&+47:15:57.80&12:18:53.806&+47:15:57.77&0.828 & --\\
12:18:54.751&+47:14:43.04&12:18:54.806&+47:14:43.19&0.841 & --\\
12:18:56.244&+47:14:51.12&12:18:56.268&+47:14:50.85&0.454 & --\\
12:18:52.056&+47:16:53.13&12:18:52.094&+47:16:53.18&0.578 & --\\
\hline
\multicolumn{6}{c}{Corrected X-3 coordinate on {\it SDSS} and {\it HST} image}\\
\hline
{\it SDSS} R.A.&{\it SDSS} Dec.&{\it HST} R.A.&{\it HST} Dec.\\
\hline
12:18:57.859&+47:16:07.44&12:18:57.902&+47:16:07.62\\
\hline
\end{tabular}
\label{T:T5}
\end{minipage}
\end{table*}

\begin{table*}
\centering
\caption{The log of {\it HST}/ACS observations.}
\begin{tabular}{cccccc}
\hline\hline
Filter & ObsID & Date & Exp. \\
& & & (ks)\\
\hline
ACS/F606W & j96h39020 & 2005-03-13 & 1.01 \\
\hline
ACS/F435W & jb1f98q8q & 2009-12-03 & 0.36 \\
ACS/F555W & jb1f98010 & 2009-12-03 & 0.98 \\
ACS/F814W & jb1f98q5q & 2009-12-03 & 0.38 \\
\hline
ACS/F435W & jb1f89eoq & 2009-12-14 & 0.36 \\
ACS/F555W & jb1f89010 & 2009-12-14 & 0.98 \\
ACS/F814W & jb1f89elq & 2009-12-14 & 0.38 \\
\hline
\end{tabular}
\label{T:T6}
\end{table*}

\begin{table*}
\centering
\caption{The dereddened magnitude and color values of optical candidates of X-3 obtained from {\it HST}/ACS data. The F435W, F555W and F814W filter correspond to {\it B}, {\it V}, {\it I} in johnson {\it BVI} system, respectively.}
\begin{tabular}{ccccccc}
\hline\hline
Date & Filter & \multicolumn{2}{c} {VEGAmag} & \multicolumn{2}{c} {Johnson Mag} \\
&&&\\
\hline
& & 1 & 2 & 1  & 2\\
\hline
2005-03-13 &F606W & $23.22\pm 0.02$ & $23.26\pm 0.02$ & - & - & \\
 & &  & \\

& 			F435W & $23.09\pm 0.04$ & $22.97\pm 0.03$ & $23.13\pm 0.04$ & $23.02\pm 0.03$ \\
2009-12-03 &F555W & $23.07\pm 0.03$ & $23.03\pm 0.03$ & $23.02\pm 0.03$ & $23.00\pm 0.03$\\
& 			F814W & $22.74\pm 0.03$ & $23.16\pm 0.04$ & $22.73\pm 0.03$ & $23.15\pm 0.04$\\ 
& $(B-V)_{0}$ &  &  & 0.11 & 0.02\\ 
& M$_{V}$ & & &-6.41 &-6.43 \\
 & & & \\
&			F435W & $23.15\pm 0.04$ & $22.90\pm 0.03$ & $23.17\pm 0.04$ & $22.96\pm 0.03$\\
2009-12-14 &F555W & $23.06\pm 0.03$ & $23.11\pm 0.03$ & $23.01\pm 0.03$ & $23.09\pm 0.03$\\
&			F814W & $22.71\pm 0.03$ & $23.31\pm 0.04$ & $22.70\pm 0.03$ & $23.30\pm 0.04$\\
& $(B-V)_{0}$ & &  & 0.16 & -0.13 \\ 
& M$_{V}$ &  & & -6.42 & -6.34 \\
& & & \\
\hline
\end{tabular}
\label{T:T7}
\end{table*}

\bsp	
\label{lastpage}
\end{document}